\documentclass[10pt, conference, letterpaper]{IEEEtran}
\usepackage[utf8]{inputenc}
\usepackage[english]{babel}
\usepackage[pdftex]{graphicx}
\usepackage{cite}
\usepackage{amsmath}
\usepackage{amssymb}
\usepackage{caption}
\usepackage{subcaption}
\usepackage{stfloats}

\usepackage{tabularx}
\usepackage{multicol}
\usepackage{epstopdf}
\usepackage{enumitem}
\usepackage{url}
\graphicspath{{Pics/}}
\DeclareGraphicsExtensions{.eps,.pdf,.png,.jpg}

\IEEEoverridecommandlockouts

\usepackage{tikz}
\usepackage{textcomp}
\usepackage{hyperref}

\newcommand\copyrighttext{%
\footnotesize \textcopyright 2020 IEEE. Personal use of this material is permitted.
Permission from IEEE must be obtained for all other uses, in any current or future
media, including reprinting/republishing this material for advertising or promotional
purposes, creating new collective works, for resale or redistribution to servers or
lists, or reuse of any copyrighted component of this work in other works.
}
\newcommand\copyrightnotice{%
\begin{tikzpicture}[remember picture,overlay]
	\node[anchor=south,yshift=10pt] at (current page.south) {\fbox{\parbox{\dimexpr\textwidth-\fboxsep-\fboxrule\relax}{\copyrighttext}}};
\end{tikzpicture}%
}

\begin{document}
\bstctlcite{IEEEexample:BSTcontrol}

\title{SDR-based Testbed for Real-time CQI Prediction for URLLC}

\author{
	Kirill Glinskiy\IEEEauthorrefmark{1}\IEEEauthorrefmark{3},  Aleksey Kureev\IEEEauthorrefmark{1}\IEEEauthorrefmark{2},
	Evgeny Khorov\IEEEauthorrefmark{1}\IEEEauthorrefmark{2}\\
	\IEEEauthorblockN{\IEEEauthorrefmark{1} Institute for Information Transmission Problems, Russian Academy of Sciences, Moscow, Russia \\
		\IEEEauthorrefmark{2} National Research University Higher School of Economics, Moscow, Russia \\
		\IEEEauthorrefmark{3} Moscow Institute of  Physics and Technology, Moscow, Russia}	
	Email: \{khorov, kureev, glinsky\}@wireless.iitp.ru\\
	
	\vspace{-1.5em}}

\maketitle 
\copyrightnotice

\begin{abstract}
	Ultra-reliable Low-Latency Communication (URLLC) is a key feature of 5G systems. The quality of service (QoS) requirements imposed by URLLC are less than 10ms delay and less than $10^{-5}$ packet loss rate (PLR). To satisfy such strict requirements with minimal channel resource consumption, the devices need to accurately predict the channel quality and select Modulation and Coding Scheme (MCS) for URLLC in a proper way.
	This paper presents a novel real-time channel prediction system based on Software-Defined Radio that uses a neural network. The paper also describes and shares an open channel measurement dataset that can be used to compare various channel prediction approaches in different mobility scenarios in future research on URLLC\footnote{Support from the Basic Research Program of the National
		Research University Higher School of Economics is gratefully acknowledged.}.
\end{abstract}

\section{Introduction} 
With the advent of the 5G cellular systems, significant attention has been dedicated to the Ultra-reliable Low-Latency Communications (URLLC), which aims to provide millisecond-level latencies and packet loss ratios as small as $10^{-5}$\cite{3GPP1}.
To reliably transmit data with strict QoS requirements and save the channel resources, the network has to adjust the MCS according to channel quality. Cellular networks perform this adjustment based on Channel Quality Indicator (CQI) reports from devices. The CQI is inferred from signal to interference-and-noise ratio (SINR) measurements for specific types of pilot signals, such as Cell Reference Signal (CRS)\cite{3GPP}.

Obtaining CQI reports too frequent consumes much channel resource. Thus, since the channel quality changes rapidly, the channel estimation becomes outdated and obsolete at the time of scheduled transmission. Therefore, a candidate solution to this problem is based on the prediction of the channel quality changes with machine learning (ML)  or statistical approaches to solve the problem\cite{Belogaev2019}\cite{Abdulhasan2014}.

Naturally, these approaches, especially deep learning models, such as Neural Networks (NN), require a significant amount of training data in varied scenarios, preferably as realistic as possible, i.e., experimental. However, most developed testbeds for channel measurements\cite{Lerch2015}\cite{Rupasinghe2015} are unsuitable for this purpose because they either are complex, limiting their use in mobility scenarios, or cannot perform high-granularity measurements, necessary for fast-changing channels.
For the considered problem, the testbed shall satisfy the following requirements:

\begin{itemize}
	\item The testbed shall be capable of high granularity of channel estimation measurements for quickly changing channels. As the CQI is calculated and can be reported to the cell as often as one subframe, which spans 1 ms, the necessary granularity is 1 ms.
	\item The testbed has to be reconfigurable to function in different bandwidths and different frequency bands.
	\item The testbed shall not emit energy in the licensed spectrum.
	\item The testbed must be compatible with commercial off-the-shelf devices used by network operators so it can be evaluated in the real-life network use case.
\end{itemize}

In this paper, we present an SDR-based testbed that meets the aforementioned requirements while being compact and portable. We also share an open data set of channel measurements obtained with the presented testbed.

\section{Demo description}
We implement and evaluate our CQI measurement testbed based on the software-defined radio (SDR), namely LIME SDR. 
We use LimeSuite for data acquisition from the SDR hardware and perform data processing using a modified srsLTE\cite{srslte} framework. srsLTE provides us with the RX pipeline on the UE. We implement two extensions to the framework: (i) real-time CQI and SINR logging on the UE based on the downlink CRS signals, and (ii) a real-time neural network for future channel quality prediction. Data processing is done on a PC, which is connected to the LIME SDR via a USB3 data link.

\begin{figure}[t]
	\centering
	\includegraphics[width=0.8\linewidth]{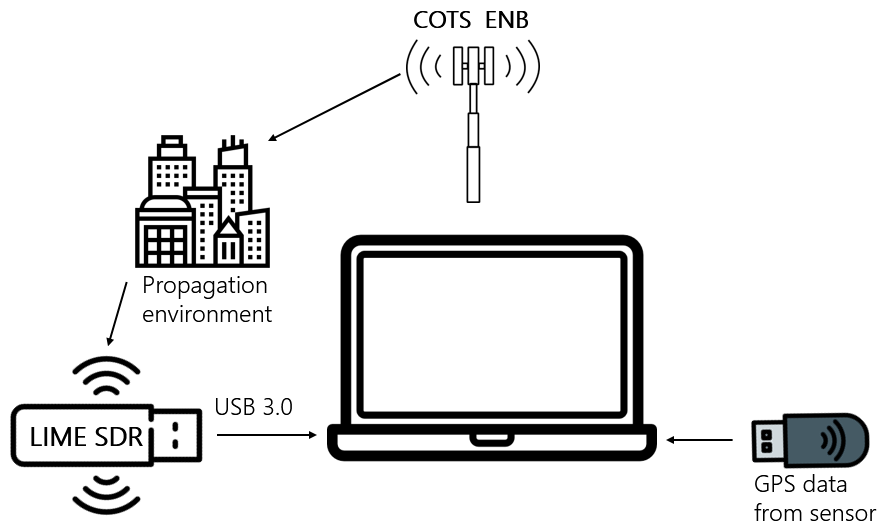}
	\caption{Testbed and experiment scheme.}
	\label{fig:testbed}
\end{figure}

We use the common modules available from srsLTE for synchronization, a cell identifier, and CRS positions search. We log the measured SNR values and input them into SINR to the CQI pipeline, which is a part of the user equipment (UE) uplink pipeline. However, we do not entirely implement the uplink system, as our testbed has no TX capabilities and loads the processed CQI into the neural network predictor. 
During a live demo, we perform the channel data collection from the nearby cells deployed by the cellular operator, as well as run the channel prediction on the incoming data. Fig. \ref{fig:testbed} shows the scheme of our demo. With different configurations, the testbed can work with existing 4G and emerging 5G deployments.

\section{Neural network predictor}

\begin{figure}[t]
	\centering
	\includegraphics[width=0.8\linewidth]{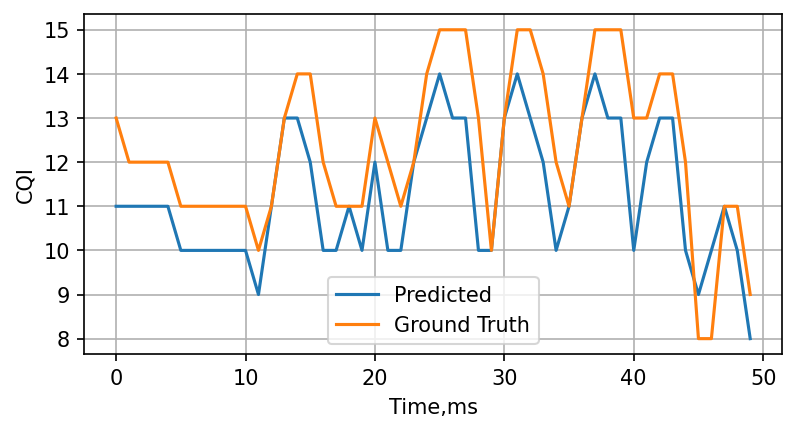}
	\centering
	\caption{Example of CRNN predictions across one RB.} 
	\label{fig:preds}
\end{figure}

We implement a neural network channel quality predictor on the testbed. The developed neural network is based on Convolutional-Recurrent Neural Network (CRNN).  The neural network input is the previous values of channel quality indicator across all resource blocks (RB) in the form of a time-frequency grid, and we predict the CQI values across all RB for the next moment in the future. We train the neural network using an asymmetric loss function on the train split of the gathered CQI dataset. During the testbed demo, the network runs in real-time. Fig. \ref{fig:preds} shows an example of the predicted and measured values comparison.  Note that for this prediction, the loss function was adjusted to provide a conservative estimate for highly reliable transmissions.

In addition, while the channel quality prediction based on CQI for scheduling transmissions generally takes place on the base station, we deploy the network on the UE at the moment to simplify the testbed architecture.
\section{Gathered dataset} 
We gather and release a novel channel quality dataset in multiple mobility and stationary scenarios. The dataset consists of CQI and SINR across all RB in the selected bandwidth. The measurements included in the dataset consist of the following scenarios:

\begin{itemize}
	\item{High-speed train channel with average speed of 80 km/h.}
	\item{In-Vehicle scenario with average speed of 60 km/h.}
	\item{Pedestrian scenario with average speed of 3 km/h.}
\end{itemize}

We record the channel with the channel width of 10MHz, which equates to 50 RB.  Our dataset is available at \url{http://wireless.iitp.ru/dataset-for-cellular-networks}, and we plan to extend it with more possible scenarios in the future.  The structure of the dataset is as follows: each scenario is represented with a time-frequency grid in the 2D tensor form of SINR values. 

\begin{figure}[h]
	\begin{subfigure}{1\linewidth}
		\centering
		\includegraphics[width=0.8\linewidth]{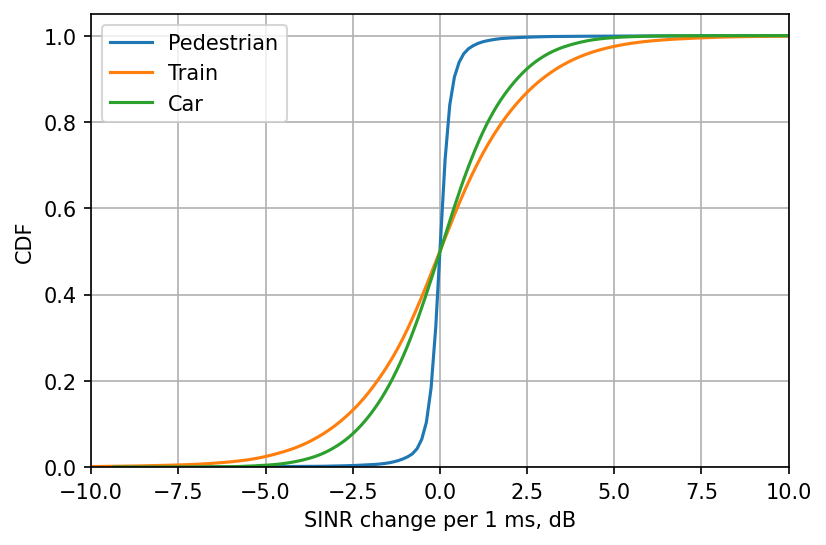}
	\end{subfigure}
	\caption{CDF plot of SINR change per 1 ms across multiple scenarios.}
	\label{fig:time}
\end{figure}

Here we perform minimal exploratory analysis on the collected dataset by plotting the CDF plot of observed SINR change per 1 ms on Fig. \ref{fig:time}. Here, we can observe that the distribuitons are similar in form, likely due to having similar line-of-sight conditions. However, due to higher vehicle speed and therefore Doppler frequency, the train and car scenarios exhibit much faster SINR changes. 	

In this paper, we have outlined the requirements for an efficient and portable channel measurement testbed. We developed an SDR-based testbed that satisfies these requirements. Using this testbed, we obtained channel measurements in multiple cellular network usage scenarios. Finally, we implemented and trained a novel CQI prediction algorithm based on CRNN using the gathered dataset. We release the gathered dataset to facilitate further research in this direction.

\bibliographystyle{IEEEtran}
\bibliography{biblio}

\begin{thebibliography}{1}
\providecommand{\url}[1]{#1}
\csname url@samestyle\endcsname
\providecommand{\newblock}{\relax}
\providecommand{\bibinfo}[2]{#2}
\providecommand{\BIBentrySTDinterwordspacing}{\spaceskip=0pt\relax}
\providecommand{\BIBentryALTinterwordstretchfactor}{4}
\providecommand{\BIBentryALTinterwordspacing}{\spaceskip=\fontdimen2\font plus
\BIBentryALTinterwordstretchfactor\fontdimen3\font minus
  \fontdimen4\font\relax}
\providecommand{\BIBforeignlanguage}[2]{{%
\expandafter\ifx\csname l@#1\endcsname\relax
\typeout{** WARNING: IEEEtran.bst: No hyphenation pattern has been}%
\typeout{** loaded for the language `#1'. Using the pattern for}%
\typeout{** the default language instead.}%
\else
\language=\csname l@#1\endcsname
\fi
#2}}
\providecommand{\BIBdecl}{\relax}
\BIBdecl

\bibitem{3GPP1}
3GPP, ``{Study on physical layer enhancements for NR ultra-reliable and low
  latency case (URLLC)},'' 2 2018, v10.7.0.

\bibitem{3GPP}
3GPP, ``{Evolved Universal Terrestrial Radio Access (E-UTRA)},'' 2 2013,
  v10.7.0.

\bibitem{Belogaev2019}
A.~Belogaev, E.~Khorov, A.~Krasilov, D.~Shmelkin, and S.~Tang, ``Conservative
  link adaptation for ultra reliable low latency communications,'' in
  \emph{2019 {IEEE} International Black Sea Conference on Communications and
  Networking ({BlackSeaCom})}.\hskip 1em plus 0.5em minus 0.4em\relax {IEEE},
  jun 2019.

\bibitem{Abdulhasan2014}
M.~Q. Abdulhasan, M.~I. Salman, C.~K. Ng, N.~K. Noordin, S.~J. Hashim, and
  F.~B. Hashim, ``A channel quality indicator ({CQI}) prediction scheme using
  feed forward neural network ({FF}-{NN}) technique for {MU}-{MIMO} {LTE}
  system,'' in \emph{2014 {IEEE} 2nd International Symposium on
  Telecommunication Technologies ({ISTT})}.\hskip 1em plus 0.5em minus
  0.4em\relax {IEEE}, nov 2014.

\bibitem{Lerch2015}
M.~Lerch, ``Experimental comparison of fast-fading channel interpolation
  methods for the {LTE} uplink,'' in \emph{2015 57th International Symposium
  {ELMAR} ({ELMAR})}.\hskip 1em plus 0.5em minus 0.4em\relax {IEEE}, sep 2015.

\bibitem{Rupasinghe2015}
N.~Rupasinghe and I.~Guvenc, ``Capturing, recording, and analyzing {LTE}
  signals using {USRPs} and {LabVIEW},'' in \emph{{SoutheastCon} 2015}.\hskip
  1em plus 0.5em minus 0.4em\relax {IEEE}, apr 2015.

\bibitem{srslte}
I.~Gomez-Miguelez, A.~Garcia-Saavedra, P.~D. Sutton, P.~Serrano, C.~Cano, and
  D.~J. Leith, ``{srsLTE: an open-source platform for LTE evolution and
  experimentation},'' in \emph{Proceedings of the Tenth ACM International
  Workshop on Wireless Network Testbeds, Experimental Evaluation, and
  Characterization}.\hskip 1em plus 0.5em minus 0.4em\relax ACM, 2016, pp.
  25--32.

\end{thebibliography}
\end{document}